\documentclass[aip,jcp,preprint,twocolumn]{revtex4-1}

\usepackage{amsmath}
\usepackage{amsfonts}
\usepackage{amssymb}
\usepackage{graphicx}

\begin{document}

\title{Analytic First and Second Derivatives for the 
       Fragment Molecular Orbital Method Combined with 
       Molecular Mechanics}

\newpage
\author{Hiroya Nakata}
\email{hiroya.nakata.gt@kyocera.jp}
\affiliation{Kyocera Corporation, Research Institute for Advanced Materials 
    and Devices, 3-5-3 Hikaridai Seika-cho Soraku-gun Kyoto 619-0237, Japan.}
\author{Dmitri G. Fedorov}
\email{d.g.fedorov@aist.go.jp}
\affiliation{Research Center for Computational Design of Advanced Functional 
             Materials (CD-FMat), National Institute of Advanced Industrial 
             Science and Technology (AIST), 1-1-1 Umezono, Tsukuba, 
             Ibaraki 305-8568, Japan}

\begin{abstract}
  Analytic first and second derivatives of the energy are developed
  for the fragment molecular orbital method interfaced with molecular
  mechanics in the electrostatic embedding scheme at the level of Hartree-Fock
  and density functional theory.
  The importance of the orbital response terms is demonstrated.
  The role of the electrostatic embedding upon molecular vibrations is
  analyzed, comparing force field and quantum-mechanical treatments
  for an ionic liquid and a solvated protein.
  The method is applied for 100 protein conformations sampled in MD to take into 
  account the complexity of a flexible protein structure in solution,
  and a good agreement to experimental data is obtained: frequencies from
  an experimental IR spectrum are reproduced within 17 cm$^{-1}$.
\end{abstract}


\maketitle



\section{Introduction}

Many phenomena in biochemistry and material science involve large
molecular systems, that are hard to compute using quantum-mechanical
(QM) methods for all atoms. Geometry optimizations and vibrational
frequency calculations are especially time consuming.
Harmonic frequencies can be obtained by diagonalizing the Hessian matrix,
constructed from the second derivatives of the energy with
respect to nuclear coordinates \cite{Pulay01}, usually performed at 
a minimum from a geometry optimization requiring analytic first derivatives.

To reduce the cost of calculations, one can use various
hybrid QM and molecular mechanics (MM) methods,\cite{QMMM1,QMMM2,ONIOM,QMMM3}
including polarizable force fields. \cite{EFP2013,QuanPol}
It is also possible to reduce the number of degrees of freedom
(active atoms) in partial Hessian methods. \cite{partHess,mbh3}
    
QM calculations can be accelerated with linear scaling methods
\cite{linsca1,linsca2}. One route to accomplish the
scaling reduction is provided by fragmentation approaches
\cite{frgrev,MCF,XPol,pdf1,eemb2,MFCC3,PMISP,EFPprot,herbert,dc3}. 
Some of them have analytic second derivatives developed
\cite{IMCMO,GEBFHESS,SMFHESS,MTAhess,QMQM}.

The fragment molecular orbital (FMO) method\cite{FMO01,FMOrev1,FMOrev2,FMOrev3,FMOrev4}, is a 
fragmentation approach
for which analytic first \cite{FMORHF,FMODFT8} and second derivatives
\cite{FMO_HESS,FMO_HESS3} have been developed.
   FMO has been extensively used for analyzing protein-ligand interactions  \cite{FMOapp1,FMOapp2,FMOapp3} and properties of large systems \cite{MEP,MEPPCM} taking advantage of fragmentation for acceleration\cite{OMP} and analysis.
The cost of 
geometry optimizations for flexible systems is high because of many degrees of freedom.
Systems containing hundreds \cite{FMORaman} or thousands \cite{nanoring} of 
atoms have been fully optimized with ab initio 
FMO methods. 
For larger systems one can use parametrized FMO methods \cite{FMO-DFTB,DFTB_PCM} to optimize all atoms or the frozen domain (FD) \cite{FMO_FD,fddhessian} formulation of FMO to optimize an active site.

A different route is given by a combination of QM and MM, in which all atoms can be optimized.
The simpler method, the so called mechanical embedding or IMOMM, \cite{Morokuma00} has been interfaced with
FMO. \cite{FMOopt,FMOMM2} The more accurate aproach, the electronic embedding,
also known as QM/MM \cite{QMMM1,QMMM2}, has been interfaced with FMO using 
traditional non-polarizable \cite{FMOMM1} and QM-based polarizable \cite{FMOEFP_MD} force fields. FMO/IMOMM has no electronic coupling between QM 
and MM, and the analytic FMO gradient can be easily obtained.
QM/MM based FMO with non-polarizable force fields\cite{FMOMM1} has been developed for an approximate gradient neglecting response terms.\cite{FMORHF}
Both electronic and mechanical embeddings in FMO/MM have no analytic Hessians.

The objective of this work is to develop fully analytic first and second derivatives for 
the two and three-body expansions of FMO combined with MM in the electronic (QM/MM) embedding.
The accuracy of the first and second derivatives is established on a
set of representative systems, and the method is applied to study spectra of 
an ionic liquid and a protein in solution. Parallel efficiency is also reported.
In this work, only non-polarizable force fields are used.

\section{Methodology}
\subsection{Summary of FMO based QM/MM}

GAMESS \cite{GAMESS1,GAMESS3} and LAMMPS \cite{LammpsPack} packages are used for QM and MM, respectively. 
LAMMPS is converted into a library, which is linked to GAMESS into
a single executable file. QM calculations are performed using FMO.
Two input files are used, one in the GAMESS format (for QM and link atoms) and another in the LAMMPS format
(for all atoms).

\begin{figure}
          \begin{center}
            \includegraphics[clip,width=12.0cm]{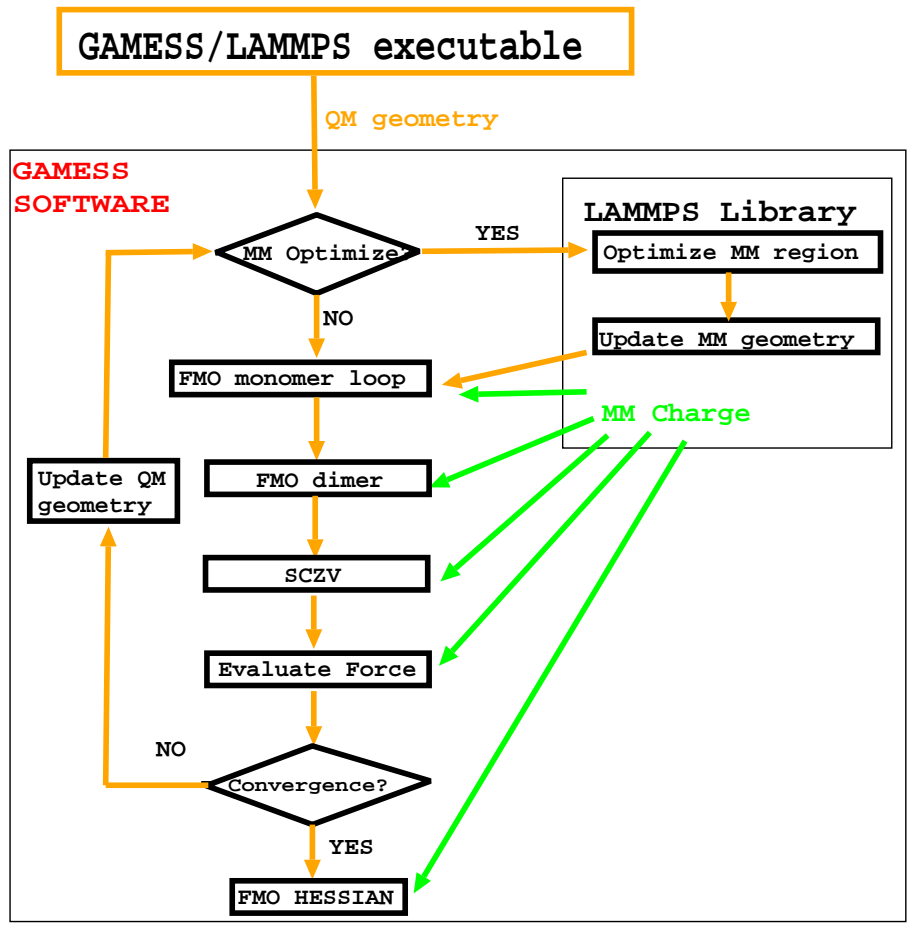} \\
          \end{center}
	      \caption{
              Schematic illustration of a geometry optimization and 
              Hessian calculation for FMO-based QM/MM.
              \label{Figure01}
		      }
	\end{figure}

The computational procedure is 
depicted in Figure~\ref{Figure01}. 
Geometry optimizations\cite{FMOopt} are performed with microiterations:
first, MM is used to optimize MM atoms, then a single optimization step is 
taken to move QM atoms, followed again by an MM optimization.

A single QM (=FMO) optimization step consists of calculating
fragments in the embedding potential (FMO monomer loop in Figure~\ref{Figure01}),
followed by embedded calculations of fragment pairs (dimers) and
self-consistent Z-vector calculations (SCZV) \cite{FMORHF} needed for analytic
gradients. SCZV is based on the coupled perturbed Hartree-Fock equations\cite{GRADBOOK}.

The embedding electrostatic potential (ESP) applied to QM consists of two parts:
standard ESP from FMO fragments \cite{ESPPTC1} and a field of point charges from MM 
atoms. In this work, periodic boundary conditions for MM are not implemented.

Regarding the naming of the method, although FMO-QM/MM was used before 
\cite{FMOMM1}, FMO/MM is the preferred notation covering both the 
mechanical embedding introduced earlier \cite{FMOopt,FMOMM2} and the electronic enbedding in this work. Alternatively, one can view the electronic embedding as a multilayer FMO \cite{MFMO}, in which the lowest layer is MM.
 
\subsection{Analytic first derivative}

The total energy for QM/MM based on FMO2 for a QM system divided into $N$ fragments is
\begin{equation}
   E  =  \sum_{I}^{N} E_{I}^{\prime}
      +  \sum_{I>J}^{N}(E_{IJ}^{\prime}-E_{I}^{\prime}-E_{J}^{\prime})
      +  \sum_{I>J}^{N}\mathrm{Tr}(\Delta\mathbf{D}^{IJ}\mathbf{V}^{IJ}) 
      \label{ENR1}%
\end{equation}
where $E_{I}^{\prime}$ and $E_{IJ}^{\prime}$ are the internal energies of 
monomer $I$ and dimer $IJ$, respectively.
$\Delta \mathbf{D}^{IJ}$ is difference between the matrices of dimer and monomer electron densities.
The QM contribution ${V}^{IJ}$ to ESP for fragment $X$ ($X= IJ$ or $I$) is
\begin{align}
 V_{ij}^{X} & =
 \sum_{K \neq X}^N 
 \left[
   \sum_{A \in K}
   \langle  
     i
     \left|
      \frac{-Z_A}{| \mathbf{r} - \mathbf{R}_A |}
     \right|
     j 
   \rangle
   +
  \sum_{k  \in K}
     2 (ij | kk)
 \right],
 \label{ESP01}
\end{align}
where $Z_{A}$ and $\mathbf{R}_A$ is the nuclear charge and coordinates of QM atom $A$,
respectively. $\mathbf{r}$ is the electron coordinate.
The internal energy of fragment $X$ for Hartree-Fock is
\begin{align}
 E_{X}^{\prime}    
  =& 
     \sum_{i   \in X}^{\mathrm{occ}} 2 \tilde{h}_{ii}^{X}
  +  \sum_{i,j \in X}^{\mathrm{occ}}
        \left[  
          2
          \left(ii|jj\right)  
        - 
          \left(ij|ij\right)
        \right]
  \notag \\
  + &   \sum_{i \in X}^{\mathrm{occ}} 2 P_{ii}^{X}
  +     E_{X}^{\mathrm{NR}} 
  +     \tilde{E}_{X}^{\mathrm{NR}} 
  +     E_{X}^{\mathrm{vdW}} 
\label{InternalE}
\end{align}
where
\begin{align}
 \tilde{h}_{ii}^{X} &= h_{ii}^{X}+\tilde{V}_{ii}^{X}
\label{hmm}
\end{align}
where $h_{ii}^{X}$, $(ii|jj)$, and $P_{ii}$  are the core Hamiltonian, two-electron integrals,
and projection operator, respectively. 
$E_{X}^{\mathrm{NR}}$ is the nuclear repulsion (NR) for QM atoms.
$\tilde{E}_{X}^{\mathrm{NR}}$ is the Coulomb interaction between QM nuclei and MM charges,
and $E_{X}^{\mathrm{vdW}}$ is the dispersion interaction between QM and MM atoms.
Throughout, Roman indices ($i,j,k$ and $l$) denote molecular orbitals (MO), and Equation (\ref{InternalE})
involves occupied (occ) orbitals only.
$\tilde{V}_{ii}^{X}$ is the Coulomb interaction between electrons in QM and 
MM point charges,
\begin{align}
     \tilde{V}_{ij}^{X} 
 & =
     \sum_{A \in \mathrm{MM}} 
     \langle  
       i
       \left|
        \frac{-Q_A}{| \mathbf{r} - \mathbf{R}_A |}
       \right|
       j 
     \rangle,
\end{align}
where $Q_A$ and $\mathbf{R}_A$ are 
the point charge and coordinates of MM atom $A$, respectively.
$\mathbf{V}^{X}$ in Equation~(\ref{ENR1}) is the electrostatic 
interaction between fragment $X$ and the other fragments.

In this work, density functional theory (DFT) combined with MM is also 
developed; for this, rather trivial modifications of equations are needed and they can be inferred from earlier publications \cite{FMODFT8,FMO_HESS3}.
Likewise, in this work, the three-body expansion of FMO (FMO3) is also
developed; for this, Equation (\ref{ENR1}) is modified to include three-body corections.This addition is straightforward and can be inferred from earlier publications
\cite{FMO_HESS3}.

The derivative of the total energy with respect 
to a QM nuclear coordinate $a$ is
\begin{align}
   \frac{\partial  E}{\partial a}  
      =& \sum_{I}^{N} 
         \frac{\partial E_{I}^{\prime}}{\partial a}
      +  \sum_{I>J}^{N}(
           \frac{\partial E_{IJ}^{\prime}}{\partial a} 
         - \frac{\partial  E_{I}^{\prime}}{\partial a}
         - \frac{\partial  E_{J}^{\prime}}{\partial a}  )
   \notag \\
      +& \sum_{I>J}^{N}
         \frac{\partial \mathrm{Tr}(\Delta\mathbf{D}^{IJ}\mathbf{V}^{IJ})}
              {\partial a},
      \label{ENR1.grad}%
\end{align}

The derivative of the last term in Equation (\ref{ENR1}) is evaluated essentially as in FMO without MM \cite{FMORHF} but using densities polarized by MM in $\Delta \mathbf{D}^{IJ}$.
The derivative of the internal energy $E_{X}^{\prime}$ is
\begin{align}
\frac{\partial E_{X}^{\prime}}{\partial a}  
           &  =    
               \sum_{i\in X}^{\mathrm{occ}} 
                 \tilde{h}_{ii}^{a,X}
           +   \sum_{ij\in X}^{\mathrm{occ}}
               \left[  
                 2 \left(ii|jj\right)^{a} 
                 - \left(ij|ij\right)^{a}
               \right]  
      \nonumber \\
         & +   \sum_{i\in X}^{\mathrm{occ}} 2P_{ii}^{a,X} 
           + R^{a,X}
           -  2\sum_{ij\in X}^{\mathrm{occ}} S^{a,X}_{ji}F^{\prime X}_{ji}
      \nonumber \\
         & +  \frac{\partial E_{X}^{\mathrm{NR}}}{\partial a}
           +   \frac{\partial \tilde{E}_{X}^{\mathrm{NR}}}{\partial a}
           +   \frac{\partial E_{X}^{\mathrm{vdW}}}{\partial a},
           \label{ENR3}
\end{align} 
where
\begin{align}
R^{a,X} &=
               -\sum_{m\in X}^{\mathrm{vir}}
               \sum_{i\in X}^{\mathrm{occ}}
               4 
               U_{mi}^{a,X}
               V_{mi}^{X}
\label{respX}
\end{align}
 $U_{mi}^{a,X}$ is the orbital response due to 
 the derivatives of the MO coefficients with respect to $a$. 
Superscript $a$ denotes derivatives with respect to $a$, e. g. in $P_{ii}^{a,X}$
The sum of $R^{a,X}$ terms
for all $X$ in Equation (\ref{respX}) according to Equation (\ref{ENR1}) can be written\cite{FMORHF}
\begin{align}
R^a
\label{ResponseTerm}
=&
\mathbf{Z}^T
\mathbf{B}^a_{0}.
\end{align}
where $\mathbf{Z}$ is the Z-vector (although $\mathbf{Z}$ is a rank 2 tensor indexed by two MOs, Equation \ref{ResponseTerm} is written as a scalar product of two supervectors, summing over both MO indices\cite{FMORHF}) obtained for FMO in the SCZV method \cite{FMORHF} solving the equations
\begin{align}
\mathbf{A}\mathbf{Z}=\mathbf{L}
\end{align}
$\mathbf{A}$ is the orbital Hessian matrix of second derivatives of the energy
with respect to MO coefficients (a rank 4 tensor). Its definition uses MO intergrals \cite{FMORHF}and it is not modified explicitly for MM.
$\mathbf{L}$ is the Lagrangian (a rank 2 tensor), which also does not have explicit MM contributions.
$\mathbf{B}^a_{0}$ is a rank 2 tensor, whose definition can be found elsewhere \cite{FMORHF}. 

The electrostatic MM contributions enter in Equation (\ref{hmm}) and as the
two final terms Equation (\ref{ENR3}). The dispersion contribution is the last term in Equation (\ref{ENR3}). In addition, electrostatic contributions enter SCZV
via the use of MM-modified one-electron Hamiltonian in Equation (\ref{hmm}) as a
contribution to $\mathbf{B}^a_{0}$ (basically, one adds $\tilde{V}_{ii}^{X}$
to the standard definition\cite{FMORHF}). 
 

\subsection{Analytic second derivative}
  
The second derivative of the total energy with respect to two QM atom
coordinates $a$ and $b$ is 
{\footnotesize
\begin{align}
   \frac{\partial^2  E}{\partial a\partial b}  
    & =  \sum_{I}^{N} 
         \frac{\partial^2 E_{I}^{\prime}}{\partial a\partial b}
      +  \sum_{I>J}^{N}(
           \frac{\partial^2 E_{IJ}^{\prime}}{\partial a\partial b} 
         - \frac{\partial^2  E_{I}^{\prime}}{\partial a\partial b}
         - \frac{\partial^2  E_{J}^{\prime}}{\partial a\partial b}  )
  \notag \\
    & +  \sum_{I>J}^{N}
         \frac{\partial^2 \mathrm{Tr}(\Delta\mathbf{D}^{IJ}\mathbf{V}^{IJ})}
              {\partial a\partial b}.
      \label{ENR1.hess}%
\end{align}
}
The second derivative of the last term is straightforward to evaluate
using the FMO Hessian formulation.\cite{FMO_HESS,FMO_HESS2}
 The second derivative of the internal energy $E_{X}^{\prime}$ is 

{\footnotesize
\begin{align}
        \frac{\partial^2 E_{X}^{\prime}}
             {\partial a \partial b}
 &  =   
        \sum_{i   \in X}^{{\mathrm{occ}}} 
        \left[  
           \tilde{h}_{ii}^{ab,X} 
         + P_{ii}^{ab,X} 
         + F_{ii}^{\prime ab,X}
        \right]
    \nonumber \\
 &  -   \sum_{i \in X}^{{\mathrm{occ}}} 
        2 S_{ii}^{ab,X} \epsilon_{ii}^{X}
    +   4 
          \sum_{i \in X}^{{\mathrm{occ}}} 
          \sum_{j \in X}^{{\mathrm{occ}}} 
          S_{ji}^{b,X}
          S_{ij}^{a,X}
          \epsilon_{ii}^{ X}
    \nonumber \\
 &  +   \sum_{m   \in X}^{{\mathrm{vir}}}  
        \sum_{i   \in X}^{{\mathrm{occ}}}  
        U_{mi}^{b,X} 
        \left[
        4  F_{im}^{\prime a,X}
    -   4  S_{mi}^{a,X}
           \epsilon_{ii}^{ X}
        \right.
 \notag \\
 &
    -   2 \sum_{j,l \in X}^{{\mathrm{occ}}} 
        \left.
        A_{jl,mi}^{X,X} 
        S_{jl}^{a,X} 
        \right]
    \nonumber \\
 &  -   \sum_{i   \in X}^{{\mathrm{occ}}}  
        \sum_{j   \in X}^{{\mathrm{occ}}}  
        S_{ij}^{b,X} 
        \left[
        2 F_{ij}^{\prime a,X}
    -   \frac{1}{2}
        \sum_{k,l \in X}^{{\mathrm{occ}}} 
        A_{ij,kl}^{X,X} 
        S_{kl}^{a,X} 
        \right]
    \nonumber \\
 &  -   \sum_{i \in X}^{{\mathrm{occ}}} 
        \sum_{j \in X}^{{\mathrm{occ}}} 
        S_{ij}^{a,X} 
        \left[
        2 F_{ij}^{\prime b, X} 
    -   \frac{1}{2}
        \sum_{k,l \in X}^{{\mathrm{occ}}} 
        A_{ij,kl}^{X,X} 
        S_{kl}^{b,X} 
        \right]
    \nonumber \\
 &  +   \frac{\partial^2 E_X^{\mathrm{NR}}}{\partial a \partial b}
    +   \frac{\partial^2 \tilde{E}_{X}^{\mathrm{NR}}}{\partial a \partial b} 
    +   \frac{\partial^2 E_{X}^{\mathrm{vdW}}}{\partial a \partial b} 
    -   \overline{U}^{ab,X},
  \label{Energyterm}
 \end{align}
}
where 
$S_{jl}^{X}$ is the orbital overlap.
The internal Fock matrix elements are
\begin{align}
     F_{ij}^{\prime,X}
    &=     
     \tilde{h}_{ij}^{X}
  +  \sum_{k \in X}^{\mathrm{occ}}
     \left[
        2\left(ij|kk\right)
     -    
         \left(ij|kk\right)
     \right]
  + 
     P_{ij}^{X}.
     \label{FockMatrix}
\end{align}
The orbital response contribution $\overline{U}^{ab,X}$ in Equation (\ref{FockMatrix})
cancels out when all monomer and dimer contributions are added. \cite{FMO_HESS}

To implement QM/MM, the standard FMO Hessian is modified taking into account 
the charge contribution via Equation (\ref{hmm}), NR and vdW terms in 
Equation (\ref{Energyterm}). In addition, CPHF equations solved for monomers and dimers
 in order to obtain orbital resposes $U^{a,X}_{mi}$ in Equation (\ref{Energyterm}) 
also include the electrostatic contribution via the use of the MM-modified one-electron Hamiltonian in Equation (\ref{hmm}). 

\subsection{Frozen domain formulation}
\begin{figure}
         \begin{center}
           \includegraphics[clip,width=12.0cm]{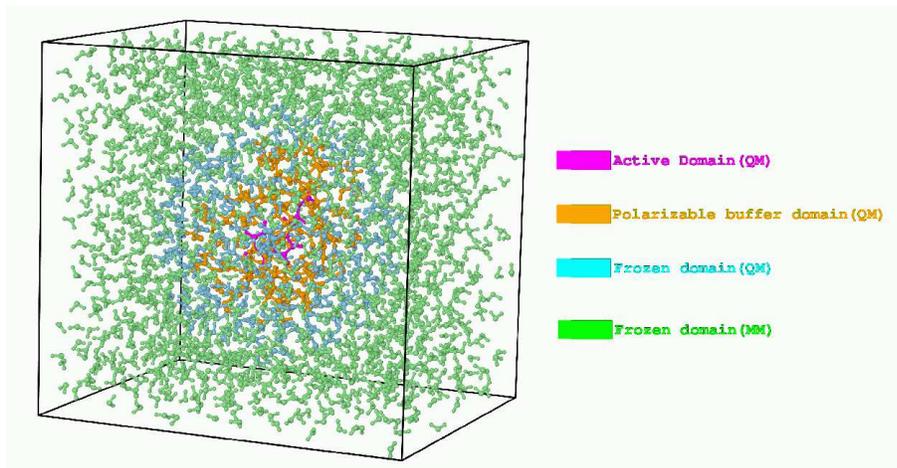} \\
         \end{center}
          \caption{
          Combination of FMO/FDD with MM.
          \label{FigureFD}
                  }
    \end{figure}

In the frozen domain formulation of FMO, all fragments are divided into
the active \textbf{A}, polarizable buffer \textbf{B} and frozen \textbf{F} 
domains (Figure~\ref{FigureFD}). Only some atoms in \textbf{A} can be optimized; all atoms in \textbf{B} and
\textbf{F} are frozen. For the initial geometry, all fragments in all domains are
calculated. During geometry optimization,
\textbf{B} is recalculated to take
into account the polarization whereas the electronic state of fragments in 
\textbf{F} is kept frozen (computed for the initial geometry only). By convention,
\textbf{B} includes \textbf{A}, because fragments in \textbf{A} are also
polarizable.

The energy of FMO/FD has an expression essentially the same as Equation (\ref{ENR1}),
except that only monomers in \textbf{A} are included in the first sum over $I$.
The most common setup is to apply the dimer (D) approximation to FD (FDD).
In FMO/FDD, the dimer sum in Equation (\ref{ENR1}) is limited to those dimers where
at least one of the two fragments ($I$ or $J$) belongs to \textbf{A}.
Energy \cite{FMO_FD}, its analytic first \cite{FMOFD2} and second \cite{fddhessian} derivatives can be calculated for FMO/FDD. 

FMO/FDD can combined with MM. The main usage of this FDD/MM approach
is to add a large polarizing environment. Most typically, in this case
only atoms in \textbf{A} can be optimized and a Hessian can be computed for
them, whereas all other atoms are frozen and contribute to the polarization.

\section{Computational details}
 The analytic first and second derivative for FMO-based QM/MM were implemented 
 into a development version of GAMESS\cite{GAMESS1} interfaced 
with LAMMPS\cite{LammpsPack}. 
FMO in GAMESS was parallelized with DDI\cite{GDDI-FMO}. MM calculations in LAMMPS were executed
sequentially.
Most simulations are performed with Hartree-Fock using the 6-31G(*) basis set 
(except where otherwise indicated).
In DFT calculations the default Lebedev grad was used.
The dispersion model D3 was employed \cite{Grimme2}. 
The second derivative of the dispersion energy is 
done by a two-point numerical differentiation. 
In comparisons to experiment calculated RHF frequencies were scaled 
\cite{SCALFAC} by the factor of 0.8953.

Molecular systems were fragmented into 1 residue and molecule per fragment
for the protein and water (ionic liquids), respectively.
As fragmentation is shifted in FMO by one carboxyl group, to make the
difference clear, fragment residues are referred to with a dash, such as Trp-6.

\begin{figure}
             \begin{center}
               \includegraphics[clip,width=12.0cm]{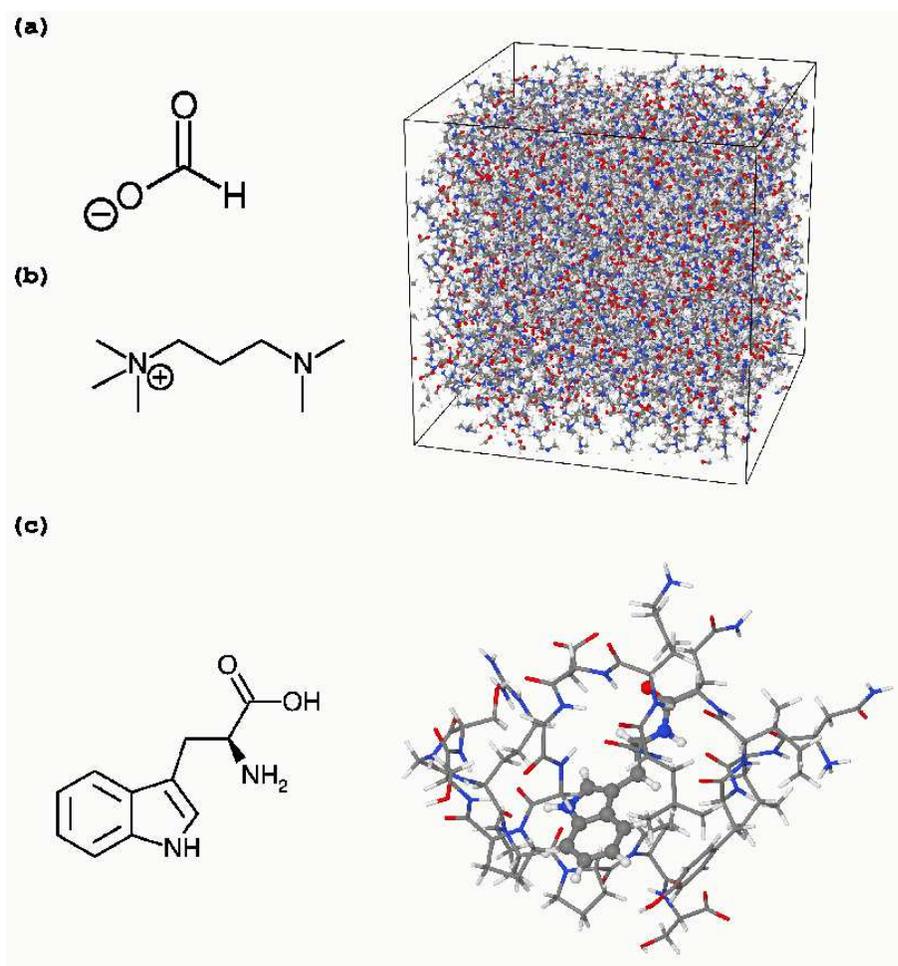} \\
             \end{center}
	      \caption{
              Molecular systems: an ionic liquid made from
              (a) formate anion and
              (b) DMEDAH cation, is shown as a box on the right.
              (c) Tryptophan amido acid and Trp-cage protein (Trp6 is
                  shown with balls and thick sticks).
              \label{Figure02}
		      }
	\end{figure}

The accuracy of the analytic energy gradient is evaluated 
by comparing it with the numerical gradient for
an ionic liquid consisting of
formate anions and dimethylethylene-diamine (DMEDAH) cations
(Figure~\ref{Figure02}-(a) and (b)).
Two anions and two cations (4 fragments) are treated with
QM and the rest is computed with MM; FDD is not used.
The ionic liquid cluster treated in QM/MM consists of 20056 atoms.
  
The effect of replacing QM with MM treatment was evaluated using FMO/FDD for 
the Trp-cage miniprotein (PDB: 1L2Y) \cite{1L2Y}
solvated in 1972 water molecules ( 6221 atoms total, Figure~\ref{Figure02}-(c)).
\textbf{A} was set to 1 fragment, Trp-6. 
\textbf{B} and \textbf{F} were constructed by including all fragments within 
6.5 and 20 \AA\ from Trp-6, respectively. \textbf{F} was computed with a 
smaller basis set STO-3G. The remaining water (outside 20 \AA) was treated in 2 ways: 
a) as a part of \textbf{F} with QM or b) with MM. 

The effect of the QM region size was evaluated for
the IR spectra of an ionic liquid and Trp-cage using FMO/FDD.
Three types of calculations were done:
a) setting one formate anion as \textbf{A} in the ionic liquid cluster,
b) setting one DMEDAH cation as \textbf{A} in the ionic liquid cluster,
c) setting Trp-6 as \textbf{A} in solvated Trp-cage.
Then, all fragments within 6.5 \AA\ in the respective system were assigned to \textbf{B}.
All fragments within the size of the QM region (6-20 \AA) were assigned
to \textbf{F}. 
Two kinds of calculations were done:
a) all fragments beyond the QM size were treated with MM (envir=MM) and
b) all fragments beyond the QM size were removed (envir=none).

In order to probe the effect of the environment upon the spectra of a
residue (Trp-6) in Trp-cage, the results for a methyl-capped Trp residue
are compared to the results for the protein
in vacuum and in solution (water).
For the capped residue, full unfragmented calculations were performed.
For the vacuum calculation, MM was not used; for the solvated protein,
MM was used for fragments beyond the QM size of 20 \AA.
FMO calculations with and without MM were done at the FDD level as described
above, i.e., Trp-6 was assigned as \textbf{A}, and all residues within
6.5 \AA\ were assigned to \textbf{B}.

The parallel efficiency was evaluated for the exact
analytic gradient and Hessians applied to solvated Trp-cage 
(FDD as described above, with the QM size of 20 \AA). 
The calculations were performed on 1 and 8 nodes 
(3.1 GHz, 128 GB RAM and 8 cores per node) connected by Gigabit.
0.966 GB of RAM per core were used per node.

For preparing the initial structure for FMO geometry optimizations
and Hessian calculations of the ionic liquid and solvated Trp-cage,
NVT molecular dynamics (MM/MD) simulations were performed for 200 ps.
Then NPT simulations were done for 1 ns, and the final geometry 
was used in FMO.
 MD simulations were performed with the time step of 0.5 fs
 with a Nose--Hoover thermostat and the velocity Verlet integrator.
 AMBER99 force field\cite{AMBER99,GAFFcite} was used in MD for all 
 molecules except that for water TIP3P was used.


 The geometry optimization for QM and MM was performed with the thresholds of 
10$^{-4}$ hartree/bohr and 10$^{-6}$ kcal/\AA, respectively.
 In simulating IR spectra, the Gaussian broadening \cite{FMORaman} was used
 with the broadening parameter of 10 cm$^{-1}$, with the exception of the
 solvated Trp-cage protein, for which
100 different structures were extracted from an NPT MD trajectory 
(one geometry every 10 ps), the geometry for each structure was optimized,
Hessian computed, and the 100 discrete IR spectra were combined into one total
IR spectrum.

\section{Results and Discussion}

\subsection{Accuracy}
\begin{figure}
          \begin{center}
            \includegraphics[clip,width=12.0cm]{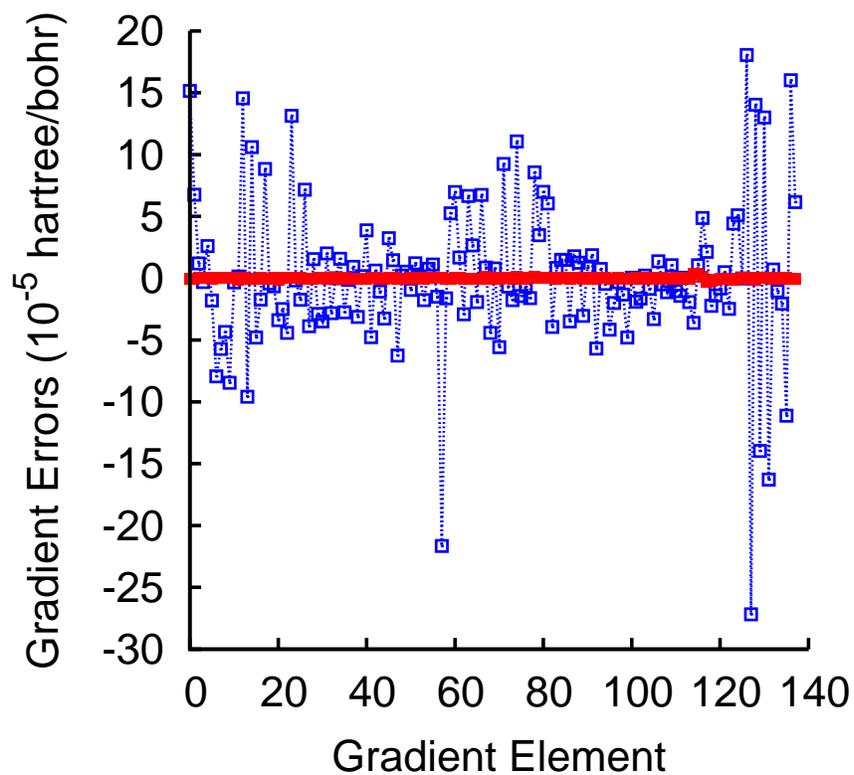} \\
          \end{center}
	      \caption{Accuracy of the exact and approximate (neglecting 
              the response term in Eq.~\ref{ResponseTerm}) analytic gradients
              for FMO-based QM/MM with respect to numerical gradient 
              for an ionic liquid, shown as
              red solid (exact) and blue dashed (approximate) lines.
              Gradient elements on the x-axis are plotted for x, y and z
              cooridnates of all atoms consequently.
              \label{Figure03}
		      }
	\end{figure}

Three kinds of gradients were computed for an ionic liquid with 4 fragments
in the QM refion: a) exact analytic, b) approximate analytic and c) numerical gradients.
The latter gradient was used as the reference to measure the accuracy of the
exact analytic gradient, and to show the importance of the response terms in Equation (~\ref{ENR3}),
neglected in the approximate gradient.
The results are shown in in Figure~\ref{Figure03} and TABLE~\ref{TABLE01}.
The exact gradient is accurate (agrees with the numerical gradient),
whereas approximate gradient neglecting response terms has an error of about $3 \times 10^{-4}$ hartree/bohr,
which is larger than the default threshold for geometry optimizations by 
a factor of 3.
   Neglecting response terms can lead to energy oscillations in
   geometry optimization and unreliable MD simulations.\cite{FMOMD1,FMORHF}

   In order to evaluate the fragmentation accuracy, FMO-based QM/MM 
calculations were compared to full QM/MM calculations (where the QM region was
not fragmented) at the DFT level. 
For this test, a water cluster was used with 9 and 365 water 
molecules in QM and MM regions, respectively. The results
   are shown in TABLE~\ref{TABLE02}. 
   The accuracy  of FMO is reasonable.
  
\begin{figure}
          \begin{center}
            \includegraphics[clip,width=12.0cm]{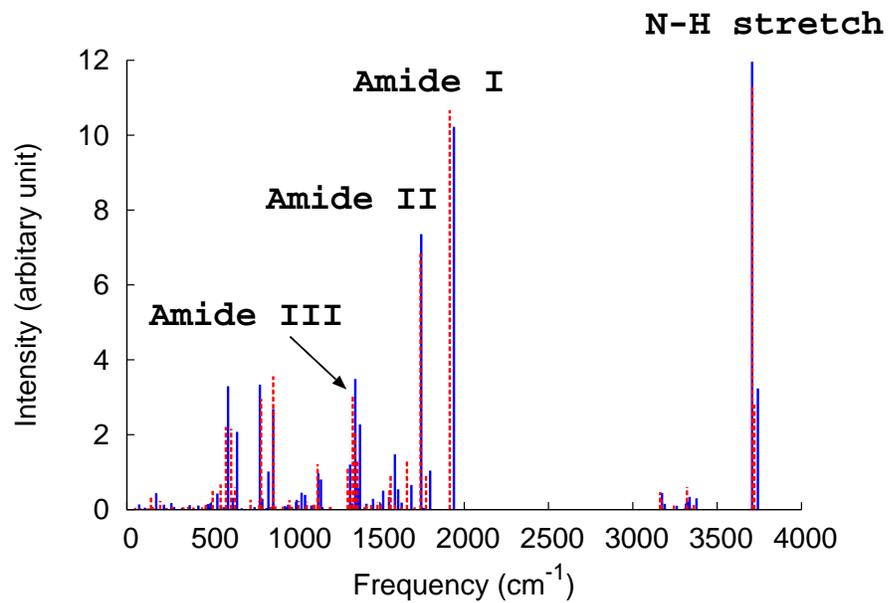} \\
          \end{center}
	      \caption{
              Effect of replacing some QM water with MM in solvated Trp-cage (envir=MM).
              Full QM and QM/MM IR spectra are shown as
              blue solid and red dashed lines, respectively.
              \label{Figure04}
		      }
	\end{figure}

   The effect of replacing QM with MM was also evaluated with the Trp-cage protein
   solvated in explicit water (the protein was treated with QM, and only the treatment of some water molecules was switched from QM to MM).
   The IR spectra are shown in Figure~\ref{Figure04}.
   There is a good agreement both in terms of frequences and intensities.
   The main peaks are listed in TABLE~\ref{TABLE02}.
   The largest deviation is 25 cm$^{-1}$.
   
\subsection{Effects of the electrostatic embedding on IR spectra} 

   IR spectra describing vibrations in a single fragment (treated as active
domain \textbf{A}) were computed for the DMEDAH-formate ionic liquid and 
Trp-cage protein, varying the QM region size. 
FMO(envir=MM) results for a range of QM sizes show the effect of replacing QM
with MM (comparison of non-polarizable vs polarizable embedding).
FMO(envir=none) results for a range of QM sizes show the effect of polarization
(how removing outward fragments changes the results).
It should be noted that both FMO(envir=MM) and FMO(envir=none) have 
exactly the same FMO/FDD setup (\textbf{A}+\textbf{B}+\textbf{F}),
dependent on the QM region size, whereas fragments beyond \textbf{F} are
either treated with MM (envir=MM) or neglected (envir=none).

\begin{figure}
          \begin{center}
            \includegraphics[clip,width=6.0cm]{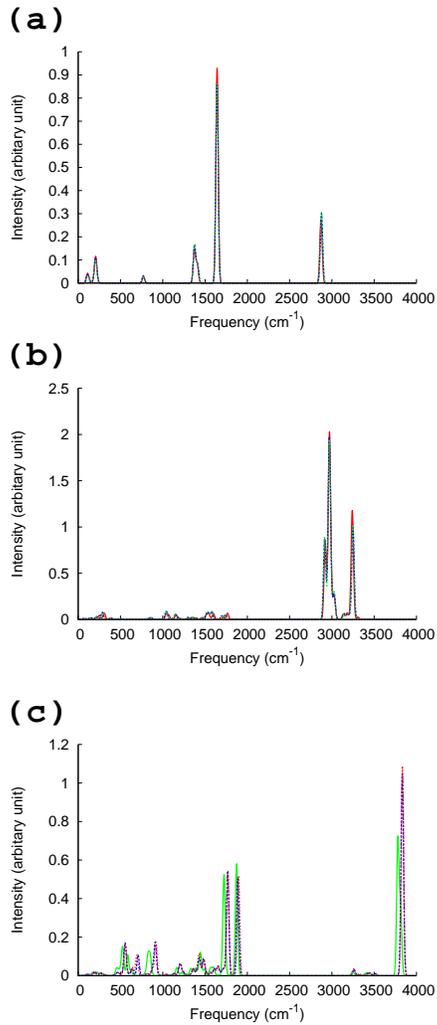} \\
          \end{center}
	      \caption{
              IR spectra of (a) one formate anion in the ionic liquid,
              (b) one DMEDAH cation the ionic liquid, and
              (c) Trp-6 in Trp-cage.
              For formate and DMEDAH, spectra are plotted for 
              the QM sizes of  6.0 \AA\ (green dashed line),
                               12.0 \AA\ (blue  dotted line),
                          and  20.0 \AA\ (red   solid  line).
              For Trp-6,
              spectra are plotted the QM sizes of  
                                8.0 \AA\ (green dashed line),
                               15.0 \AA\ (blue  dotted line),
                          and  20.0 \AA\ (red   solid  line).
              \label{Figure05}
		      }
	\end{figure}

   The effect of the QM region size upon spectra in FMO(envir=MM) is shown
   in Figure~\ref{Figure05}. Overall, a good agreement is observed,
   indicating that a non-polarizable MM describes the electrostatic embedding of the
   active fragment well compared to QM. For Trp-6 the effect is 
   slightly larger, up to about 30 cm$^{-1}$ in the frequency.

\begin{figure}
          \begin{center}
            \includegraphics[clip,width=12.0cm]{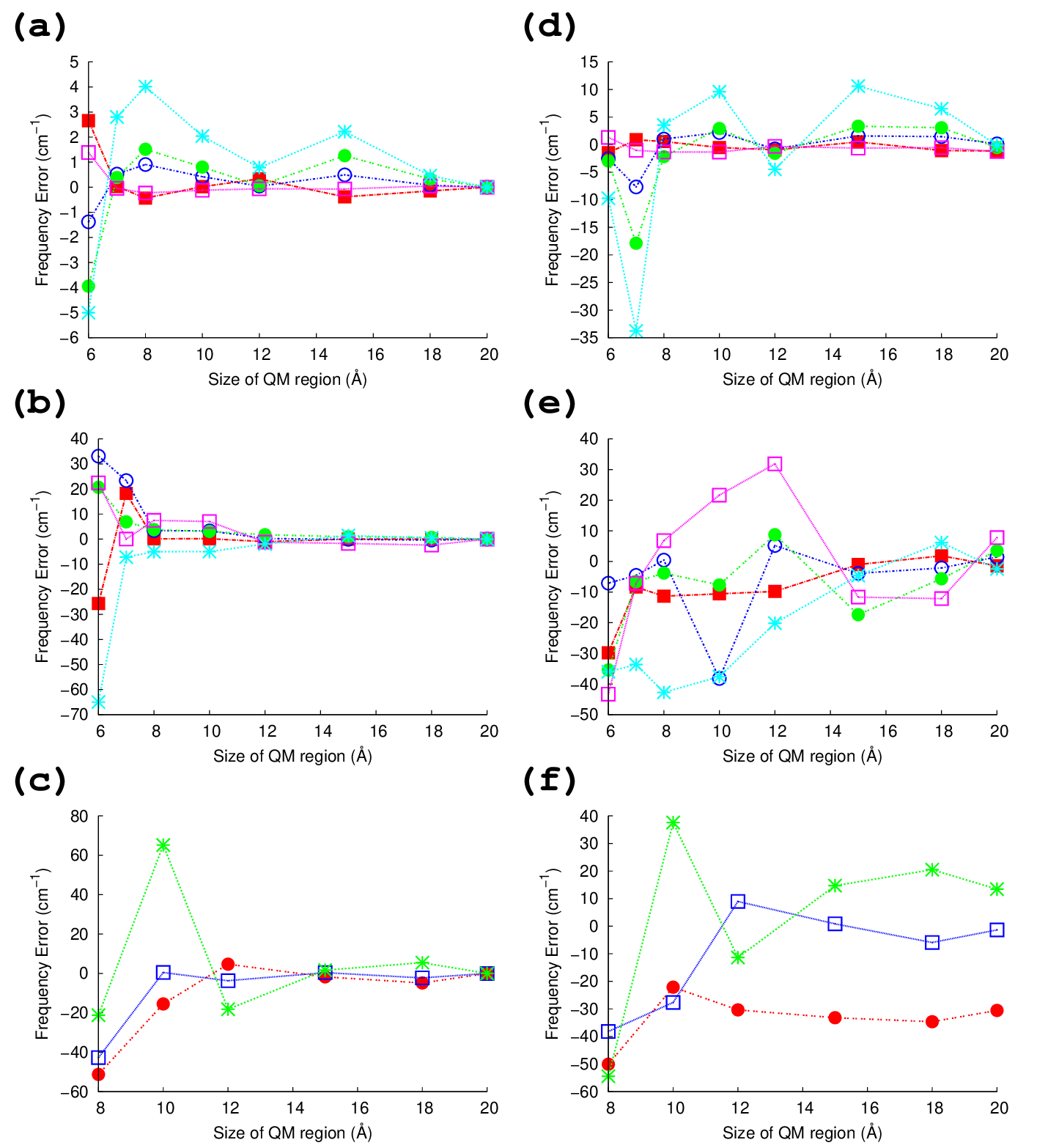} \\
          \end{center}
	      \caption{
              Deviations of the vibrational frequencies from the reference
              (20 \AA, envir=MM) for
              (a) formate (envir=MM),
              (b) DMEDAH (envir=MM),
              (c) Trp-6 (envir=MM).
              (d) formate (envir=none),
              (e) DMEDAH (envir=none),
              and (f) Trp-6 (envir=none).
              Formate and DMEDAH are computed in an ionic liquid cluster;
              Trp-6 is computed in solvated Trp-cage.
              \label{Figure06}
		      }
	\end{figure}

   The effect of the environment on frequencies
is shown in Figure~\ref{Figure06} and TABLE~\ref{TABLE03}, where
   the deviations from the reference (QM size of 20 \AA, envir=MM) are
   computed for 5 most important peaks in each system.
   Formate has small deviations not exceeding 10 cm$^{-1}$, 
   and DMEDAH and Trp-6 have deviations up to 60 cm$^{-1}$.

   For envir=MM, the results quickly converge as the QM size increases:
   for the sizes larger than 10 \AA, the deviations are less than 20 cm$^{-1}$.
   This means that the force field is adequate for such domain sizes
   and well reproduces QM results.

   For envir=none, much larger QM sizes are needed; for the ionic liquid,
   the results converge at about 20 \AA, and for Trp-6 at the largest computed
   size of 20 \AA\ there is still a deviation of about 30 cm$^{-1}$.
    The large error for Trp-6 is attributed to water molecules,
   and tryptophan is quite sensitive to the surrounding environment. 

\begin{figure}
          \begin{center}
            \includegraphics[clip,width=12.0cm]{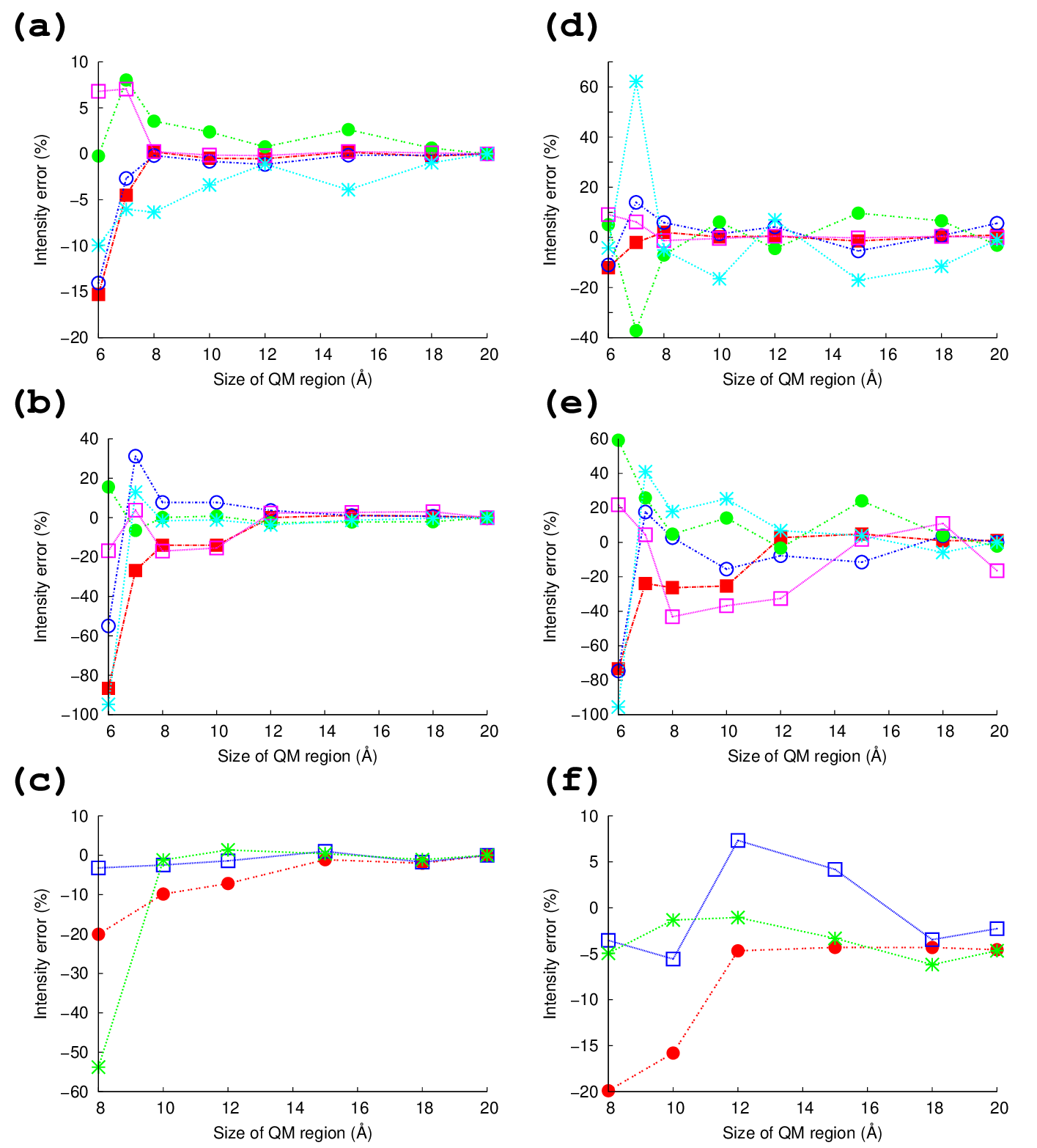} \\
          \end{center}
	      \caption{
              Deviations of the IR intensities from the reference
              (20 \AA, envir=MM) for
              (a) formate (envir=MM),
              (b) DMEDAH (envir=MM),
              (c) Trp-6 (envir=MM).
              (d) formate (envir=none),
              (e) DMEDAH (envir=none),
              and (f) Trp-6 (envir=none).
              Formate and DMEDAH are computed in an ionic liquid cluster;
              Trp-6 is computed in solvated Trp-cage.
              \label{Figure07}
		      }
	\end{figure}

   Intensities (Figure~\ref{Figure07} and TABLE~\ref{TABLE02a}), behave in a way similar to frequencies: quickly converge for
   envir=MM, and feature oscillations for envir=none.
    For the ionic liquid, envir=none has residual errors of about 16\% at 20 \AA.

   This indicates that adding environment with MM is a good way to
   improve the accuracy of the IR spectra: with the QM size of 15 \AA,
   the largest deviation is about 6 cm$^{-1}$ in frequencies.
   For 15 \AA\ in solvated Trp-cage, there are 1138 atoms in the QM region,
   which makes an application of a conventional QM/MM Hessian problematic.
   In this work, FMO/FDD is used to efficiently optimize geometry and compute Hessian
   for such QM region size,
   and relative errors in intensities are 3.9 \%.

\subsection{Effect of the protein and water environment upon vibrations}
    
   In order to investigate how the surrounding environment 
   affects the vibration frequencies in tryptophan,
   four different calculations were performed for
   (a) only the tryptophan molecule caped with methyls,
   (b) Trp-cage protein in gas phase, 
   (c) Trp-cage protein in water (at one minimum)
   (d) Trp-cage protein in water (combining 100 conformation from MD).

\begin{figure}
          \begin{center}
            \includegraphics[clip,width=12.0cm]{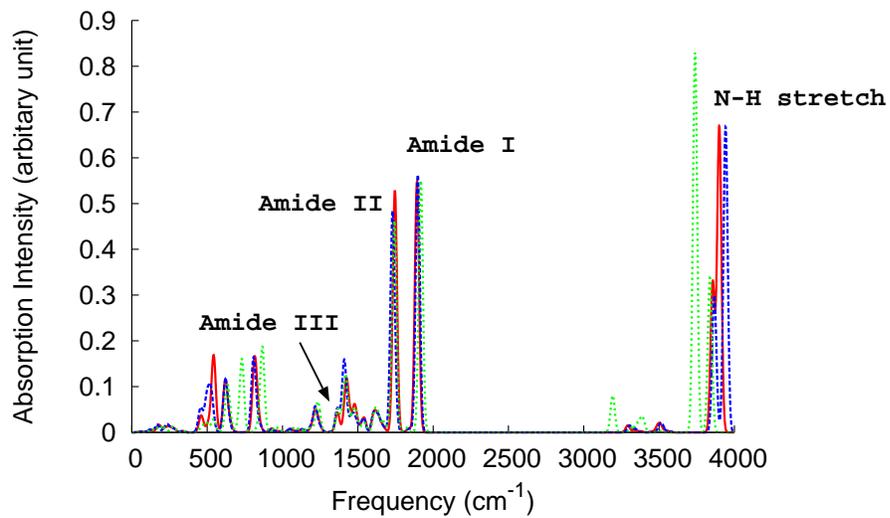} \\
          \end{center}
	      \caption{
              IR spectrum of Trp in a different environment:
              none (green dotted line), protein (blue dashed line) and
              protein+water (red solid line).
              \label{Figure08}
		      }
	\end{figure}

    The protein environment makes a substantial effect on vibrations: 
    the largest effect is for the vibration mode of N-H stretch around 3900 cm$^{-1}$ (Figure~\ref{Figure08}), and the peak of w7 is shifted by 57 cm$^{-1}$ (TABLE~\ref{TABLE02b}).
    Comparing the computed results to experiment
    without the protein environment, some frequencies differ by about
    50 cm$^{-1}$.
    Frequencies calculated for the protein in vacuum are close to solvated experiment,
    but the mode w18 differs by 29 cm$^{-1}$.
    Inclusion of water environment shifts the peak for w16 by 28 cm$^{-1}$.

\begin{figure}
          \begin{center}
            \includegraphics[clip,width=12.0cm]{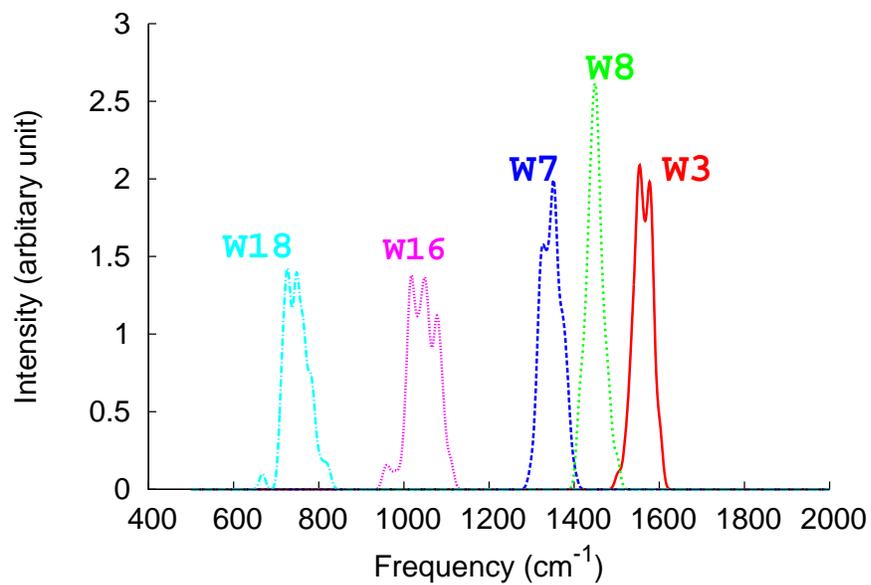} \\
          \end{center}
	      \caption{
              Computed IR spectrum for Trp-6 in solvated Trp-cage with
              important peaks w$i$ labelled.
              \label{Figure09}
		      }
	\end{figure}

    For a more reliable evaluation of spectra taking into account the existence
of multiple minima, 100 conformations were extracted from MD, and their structures were optimized. The computed combined spectrum is shown in Figure~\ref{Figure09} focusing on w$i$ peaks.
    The vibrational modes with lower frequencies are more affected by 
    the averaging, with a width of the peaks of about 150 cm$^{-1}$.
There is some structure in the peaks (i.e., each peak is a combination of
multiple peaks) that would be absent in a simple averaged broadening.
    The largest change due to averaging is about 13 cm$^{-1}$ (w3), smaller than
the effect of the electrostatic influence of the environment (TABLE~\ref{TABLE02b}).
    For the 5 peaks, the leargest deviation from experiment is for w18, 17 cm$^{-1}$.

\subsection{Parallel efficiency}
\begin{figure}
          \begin{center}
            \includegraphics[clip,width=12.0cm]{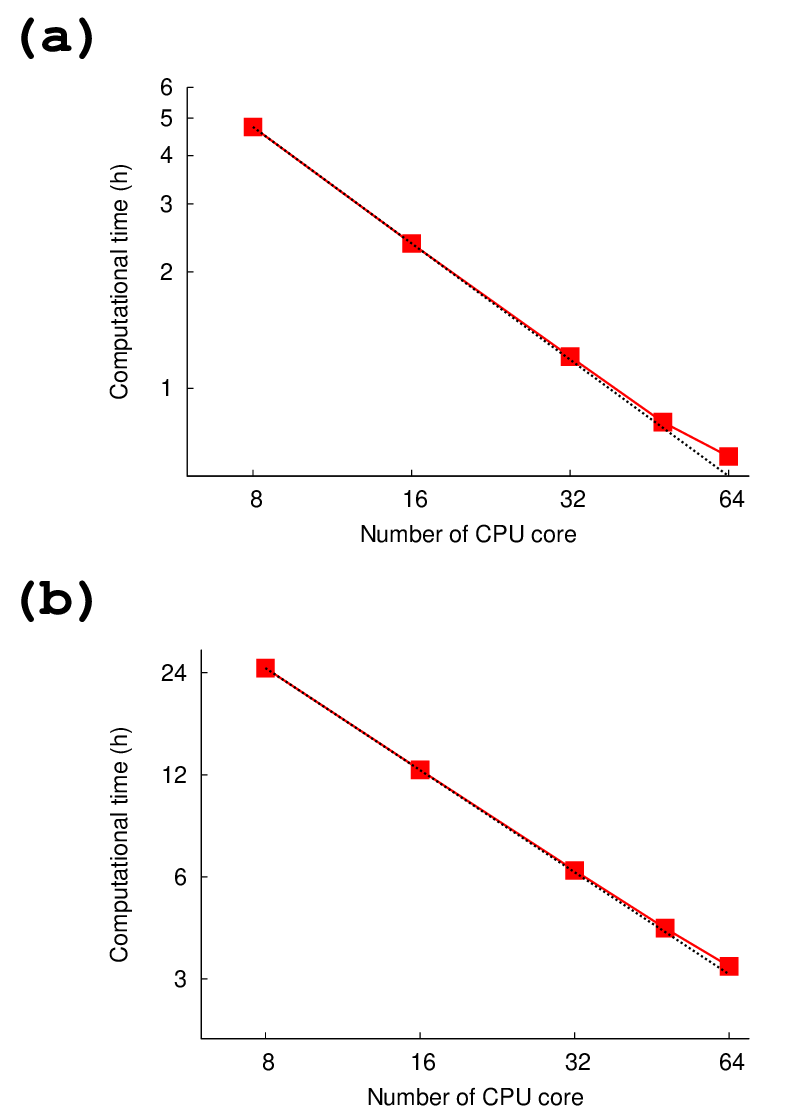} \\
          \end{center}
	      \caption{
              Wall-clock timing for FMO based QM/MM (a) gradient and (b) Hessian measured for Trp-cage solvated in explicit water: red solid and blue dashed lines show the measured and ideal results, respectively.
              \label{Figure10}
		      }
	\end{figure}

The parallel efficiency of the developed FMO/MM method is evaluated 
for both analytic gradient and Hessian using
the Trp-cage protein in explicit water with the QM size of 20 \AA,
evaluated on a PC cluster using 8-64 cores.
The timing results are shown in Figure~\ref{Figure10}.

A single point gradient calculation takes 4.74  and 0.66  hours on 8 and 64
cores, respectively, with a speedup of 7.18 on an 8-fold increase in CPU cores,
corresponding to the parallel efficiency of 88.95 \%. 
 The Hessian takes 24 and 3.26 hours on 8 and 64 cores, respectively,
 achieving teh parallel speedup of 7.36 and the parallel efficiency of 94.69 \%. 

\section{Conclusion}
  The analytic energy gradient and Hessian for the electronic embedding in 
FMO/MM have been developed and implemented into GAMESS interfaced with LAMMPS.
The importance of the response terms for obtaining
an accurate gradient needed for optimizations and MD has been demonstrated.
  The atomic types and parameters can be easily set up using tools in LAMMPS,
  making it possible to do practical applications of FMO/MM.

  It has been shown that replacing the QM treatment of some part of the 
  system by MM gives accurate IR spectra provided that the size of the QM
  region is large enough (15 \AA\ or more). On the other hand, simply
  neglecting a part of the environment results in substantial errors.
  
  The role of the environment on IR spectra of proteins has been clearly
  shown: both the protein and solvent make substantial shifts.
  The developed method is practically applicable to studying vibrations
  of a chosen part of a protein while treating a large part of it with QM
  and the rest with MM, with a possibility to take into account multiple minima
  by considering a set of conformations from MD.
  A single point gradient and a Hessian calculation take 0.66 and 3.3 hours
  on 64 CPU cores, so that these simulations are practically feasible.
  Another advantage of FMO is that less memory is typically needed than for
  unfragmented calculations, because the main memory consuming step, CPHF
  equations, are performed for individual fragments in the field of the whole system.

  Vibrations in the binding pocket of a protein binding a ligand can provide
   valuable information about the structure in solution. In addition, it is
  possible to use vibrational frequencies for estimating zero point
  energy and the vibrational contribution to the Gibbs free energy,\cite{FMO_HESS}
  which can be used for increasing the accuracy of the predictions of a
  transition state barrier in enzymes or protein-ligand binding energies.\cite{FMOsawada1}
  Although unharmonic corrections are not explicitly evaluated, they can be
  approximately corrected for using the frequency scaling.\cite{SCALFAC}

  There is a substantial interest in applications of FMO to ionic liquids
  \cite{ionl1,ionl2,ionl3} and proteins. The ability to quickly calculate IR spectra and refine transition states in enzymatic reaction \cite{enzyme} 
facilitated by the
  combination of the frozen domain FMO and MM can be useful in future applications.

\section*{ACKNOWLEDGMENT}
We thank 
the Supercomputer system ITO of 
R.I.I.T.\ at Kyushu University (Japan) 
and
and Information Technology Center at the University of Tokyo
(HPCI System Research project hp200015)
for providing computational resources. 

\bibliographystyle{aip}
\bibliography{fmomm}

\begin{thebibliography}{10}

\bibitem{Pulay01}
P.~Pulay,
\newblock Mol. Phys. {\bf 17}, 197 (1969).

\bibitem{QMMM1}
A.~Warshel and M.~Karplus,
\newblock J. Am. Chem. Soc. {\bf 94}, 5612 (1972).

\bibitem{QMMM2}
J.~A. McCammon, B.~R. Gelin, and M.~Karplus,
\newblock Nature {\bf 267}, 590 (1972).

\bibitem{ONIOM}
S.~Dapprich, I.~Kom\'aromi, K.~S. Byun, K.~Morokuma, and M.~J. Frisch,
\newblock J. Mol. Str.: THEOCHEM {\bf 461}, 1 (1999).

\bibitem{QMMM3}
K.~Schwinn, N.~Ferr\'{e}, and M.~Huix-Rotllant,
\newblock J. Chem. Theory Comput. {\bf in press}, 10.1021/acs.jctc.9b01145
  (2020).

\bibitem{EFP2013}
M.~S. Gordon, Q.~A. Smith, P.~Xu, and L.~V. Slipchenko,
\newblock Ann. Rev. Phys. Chem. {\bf 64}, 553 (2013).

\bibitem{QuanPol}
D.~S. N.~M. Thellamurege~and, F.~Cui, H.~Zhu, R.~Lai, and H.~Li,
\newblock J. Comput. Chem. {\bf 34}, 2816 (2013).

\bibitem{partHess}
H.~Li and J.~H. Jensen,
\newblock Theor. Chem. Acc. {\bf 107}, 211 (2002).

\bibitem{mbh3}
A.~Ghysels, D.~Van~Neck, V.~Van~Speybroeck, B.~R. Brooks, and M.~Waroquier,
\newblock J. Chem. Phys. {\bf 130}, 084107 (2009).

\bibitem{linsca1}
J.~Kussmann, M.~Beer, and C.~Ochsenfeld,
\newblock WIREs: Comput. Mol. Sci. {\bf 3}, 614 (2013).

\bibitem{linsca2}
A.~V. Akimov and O.~V. Prezhdo,
\newblock Chem. Rev. {\bf 115}, 5797 (2015).

\bibitem{frgrev}
M.~S. Gordon, D.~G. Fedorov, S.~R. Pruitt, and L.~V. Slipchenko,
\newblock Chem. Rev. {\bf 112}, 632 (2012).

\bibitem{MCF}
P.~Otto and J.~Ladik,
\newblock Chem. Phys. {\bf 8}, 192 (1975).

\bibitem{XPol}
J.~Gao,
\newblock J. Phys. Chem. B {\bf 101}, 657 (1997).

\bibitem{pdf1}
K.~Kiewisch, C.~R. Jacob, and J.~Visscher,
\newblock J. Chem. Theory Comput. {\bf 9}, 2425 (2013).

\bibitem{eemb2}
J.~Friedrich, H.~Yu, H.~R. Leverentz, P.~Bai, J.~I. Siepmann, and D.~G.
  Truhlar,
\newblock J. Phys. Chem. Lett. {\bf 5}, 666 (2014).

\bibitem{MFCC3}
J.~Liu, X.~Wang, J.~Z.~H. Zhang, and X.~He,
\newblock RSC Adv. {\bf 5}, 107020 (2015).

\bibitem{PMISP}
M.~Andreji\'c, U.~Ryde, R.~A. Mata, and P.~S\"oderhjelm,
\newblock Chem. Phys. Chem. {\bf 15}, 3270 (2014).

\bibitem{EFPprot}
P.~K. Gurunathan, A.~Acharya, D.~Ghosh, D.~Kosenkov, I.~Kaliman, Y.~Shao, A.~I.
  Krylov, and L.~V. Slipchenko,
\newblock J. Phys. Chem. B {\bf 120}, 6562 (2016).

\bibitem{herbert}
K.-Y. Liu and J.~M. Herbert,
\newblock J. Chem. Theory Comput. {\bf 16}, 475 (2020).

\bibitem{dc3}
N.~Komoto, T.~Yoshikawa, Y.~Nishimura, and H.~Nakai,
\newblock J. Chem. Theory Comput. {\bf 16}, 2369 (2020).

\bibitem{IMCMO}
S.~Sakai and S.~Morita,
\newblock J. Phys. Chem. A {\bf 109}, 8424 (2005).

\bibitem{GEBFHESS}
W.~Hua, T.~Fang, W.~Li, J.-G. Yu, and S.~Li,
\newblock J. Phys. Chem. A {\bf 112}, 10864 (2008).

\bibitem{SMFHESS}
M.~A. Collins,
\newblock J. Chem. Phys. {\bf 141}, 094108 (2014).

\bibitem{MTAhess}
A.~P. Rahalkar, V.~Ganesh, and S.~R. Gadre,
\newblock J. Chem. Phys. {\bf 129}, 234101 (2008).

\bibitem{QMQM}
J.~C. Howard and G.~S. Tschumper,
\newblock J. Chem. Phys. {\bf 139}, 184113 (2013).

\bibitem{FMO01}
K.~Kitaura, E.~Ikeo, T.~Asada, T.~Nakano, and M.~Uebayasi,
\newblock Chem. Phys. Lett. {\bf 313}, 701 (1999).

\bibitem{FMOrev1}
D.~G. Fedorov and K.~Kitaura,
\newblock J. Phys. Chem. A. {\bf 111}, 6904 (2007).

\bibitem{FMOrev2}
D.~G. Fedorov, T.~Nagata, and K.~Kitaura,
\newblock Phys. Chem. Chem. Phys. {\bf 14}, 7562 (2012).

\bibitem{FMOrev3}
S.~Tanaka, Y.~Mochizuki, Y.~Komeiji, Y.~Okiyama, and K.~Fukuzawa,
\newblock Phys. Chem. Chem. Phys. {\bf 16}, 10310 (2014).

\bibitem{FMOrev4}
D.~G. Fedorov,
\newblock WIREs: Comp. Mol. Sc. {\bf 7}, e1322 (2017).

\bibitem{FMORHF}
T.~Nagata, K.~Brorsen, D.~G. Fedorov, K.~Kitaura, and M.~S. Gordon,
\newblock J. Chem. Phys. {\bf 134}, 124115 (2011).

\bibitem{FMODFT8}
K.~R. Brorsen, F.~Zahariev, H.~Nakata, D.~G. Fedorov, and M.~S. Gordon,
\newblock J. Chem. Theory Comput. {\bf 10}, 5297 (2014).

\bibitem{FMO_HESS}
H.~Nakata, T.~Nagata, D.~G. Fedorov, S.~Yokojima, K.~Kitaura, and S.~Nakamura,
\newblock J. Chem. Phys. {\bf 138}, 164103 (2013).

\bibitem{FMO_HESS3}
H.~Nakata, D.~G. Fedorov, F.~Zahariev, M.~W. Schmidt, K.~Kitaura, M.~S. Gordon,
  and S.~Nakamura,
\newblock J. Chem. Phys. {\bf 142}, 164103 (2015).

\bibitem{FMOapp1}
H.~Lim, J.~Chun, X.~Jin, J.~Kim, J.~Yoon, and K.~T. No,
\newblock Sc. Rep. {\bf 9}, 16727 (2019).

\bibitem{FMOapp2}
A.~Heifetz, I.~Morao, M.~M. Babu, T.~James, M.~W.~Y. Southey, D.~G. Fedorov,
  M.~Aldeghi, M.~J. Bodkin, and A.~Townsend-Nicholson,
\newblock J. Chem. Theory Comput. {\bf 16}, 2814 (2020).

\bibitem{FMOapp3}
M.~Kusumoto, K.~Ueno-Noto, and K.~Takano,
\newblock J. Comput. Chem. {\bf 41}, 31 (2020).

\bibitem{MEP}
T.~Ishikawa,
\newblock Int. J. Quantum Chem. , e25535 (2018).

\bibitem{MEPPCM}
A.~Brekhov, V.~Mironov, and Y.~Alexeev,
\newblock J. Phys. Chem. A {\bf 123}, 6281 (2019).

\bibitem{OMP}
V.~Mironov, Y.~Alexeev, and D.~G. Fedorov,
\newblock Int. J. Quantum Chem. {\bf 119}, e25937 (2019).

\bibitem{FMORaman}
H.~Nakata, D.~G. Fedorov, S.~Yokojima, K.~Kitaura, and S.~Nakamura,
\newblock J. Chem. Theory Comput. {\bf 10}, 3689 (2014).

\bibitem{nanoring}
P.~V. Avramov, D.~G. Fedorov, P.~B. Sorokin, S.~Sakai, S.~Entani, M.~Ohtomo,
  Y.~Matsumoto, and H.~Naramoto,
\newblock J. Phys. Chem. Lett. {\bf 3}, 2003 (2012).

\bibitem{FMO-DFTB}
Y.~Nishimoto, D.~G. Fedorov, and S.~Irle,
\newblock J. Chem. Theory Comput. {\bf 10}, 4801 (2014).

\bibitem{DFTB_PCM}
Y.~Nishimoto and D.~G. Fedorov,
\newblock Phys. Chem. Chem. Phys. {\bf 18}, 22047 (2016).

\bibitem{FMO_FD}
D.~G. Fedorov, Y.~Alexeev, and K.~Kitaura,
\newblock J. Phys. Chem. Lett. {\bf 2}, 282 (2011).

\bibitem{fddhessian}
H.~Nakata, D.~G. Fedorov, T.~Nagata, K.~Kitaura, and S.~Nakamura,
\newblock J. Chem. Theory Comput. {\bf 11}, 3053 (2015).

\bibitem{Morokuma00}
T.~Matsubara, F.~Maseras, N.~Koga, and K.~Morokuma,
\newblock J. Phys. Chem. {\bf 100}, 2573 (1996).

\bibitem{FMOopt}
D.~G. Fedorov, T.~Ishida, M.~Uebayasi, and K.~Kitaura,
\newblock J. Phys. Chem. A {\bf 111}, 2722 (2007).

\bibitem{FMOMM2}
D.~G. Fedorov, N.~Asada, I.~Nakanishi, and K.~Kitaura,
\newblock Acc. Chem. Res. {\bf 47}, 2846 (2014).

\bibitem{FMOMM1}
T.~Okamoto, T.~Ishikawa, Y.~Koyano, N.~Yamamoto, K.~Kuwata, and M.~Nagaoka,
\newblock Bull. Chem. Soc. Japan {\bf 86}, 210 (2013).

\bibitem{FMOEFP_MD}
T.~Nagata, D.~G. Fedorov, and K.~Kitaura,
\newblock Theor. Chem. Acc. {\bf 131}, 1136 (2012).

\bibitem{GAMESS1}
M.~W. Schmidt, K.~K. Baldridge, J.~A. Boatz, S.~T. Elbert, M.~S. Gordon, J.~H.
  Jensen, S.~Koseki, N.~Matsunaga, K.~A. Nguyen, S.~Su, T.~L. Windus,
  M.~Dupuis, and J.~A. Montgomery,
\newblock J. Comput. Chem. {\bf 14}, 1347 (1993).

\bibitem{GAMESS3}
G.~M.~J. Barca, C.~Bertoni, L.~Carrington, D.~Datta, N.~De~Silva, J.~E.
  Deustua, D.~G. Fedorov, J.~R. Gour, A.~O. Gunina, E.~Guidez, T.~Harville,
  S.~Irle, J.~Ivanic, K.~Kowalski, S.~S. Leang, H.~Li, W.~Li, J.~J. Lutz,
  I.~Magoulas, J.~Mato, V.~Mironov, H.~Nakata, B.~Q. Pham, P.~Piecuch,
  D.~Poole, S.~R. Pruitt, A.~P. Rendell, L.~B. Roskop, K.~Ruedenberg,
  T.~Sattasathuchana, M.~W. Schmidt, J.~Shen, L.~Slipchenko, M.~Sosonkina,
  V.~Sundriyal, A.~Tiwari, J.~L. Galvez~Vallejo, B.~Westheimer, M.~Włoch,
  P.~Xu, F.~Zahariev, and M.~S. Gordon,
\newblock J. Chem. Phys. {\bf 152}, 154102 (2020).

\bibitem{LammpsPack}
S.~Plimpton,
\newblock Journal of computational physics {\bf 117}, 1 (1995).

\bibitem{GRADBOOK}
Y.~Yamaguchi, H.~F. Schaefer~III, Y.~Osamura, and J.~Goddard,
\newblock {\em A New Dimension to Quantum Chemistry: Analytical Derivative
  Methods in Ab Initio Molecular Electronic Structure Theory},
\newblock Oxford University Press, New York, 1994.

\bibitem{ESPPTC1}
T.~Nakano, T.~Kaminuma, T.~Sato, K.~Fukuzawa, Y.~Akiyama, M.~Uebayasi, and
  K.~Kitaura,
\newblock Chem. Phys. Lett. {\bf 351}, 475 (2002).

\bibitem{MFMO}
D.~G. Fedorov, T.~Ishida, and K.~Kitaura,
\newblock J. Phys. Chem. A {\bf 109}, 2638 (2005).

\bibitem{FMO_HESS2}
H.~Nakata, D.~G. Fedorov, S.~Yokojima, K.~Kitaura, and S.~Nakamura,
\newblock Chem. Phys. Lett. {\bf 603}, 67 (2014).

\bibitem{FMOFD2}
H.~Nakata, D.~G. Fedorov, T.~Nagata, K.~Kitaura, and S.~Nakamura,
\newblock J. Chem. Theory Comput. {\bf 11}, 3053 (2015).

\bibitem{GDDI-FMO}
D.~G. Fedorov, R.~M. Olson, K.~Kitaura, M.~S. Gordon, and S.~Koseki,
\newblock J. Comput. Chem. {\bf 25}, 872 (2004).

\bibitem{Grimme2}
S.~Grimme, J.~Antony, S.~Ehrlich, and H.~Krieg,
\newblock J. Chem. Phys. {\bf 132}, 154104 (2010).

\bibitem{SCALFAC}
A.~P. Scott and L.~Radom,
\newblock J. Phys. Chem. {\bf 100}, 16502 (1996).

\bibitem{1L2Y}
J.~W. Neidigh, R.~M. Fesinmeyer, and N.~H. Andersen,
\newblock Nat. Struct. Biol. {\bf 9}, 425 (2002).

\bibitem{AMBER99}
J.~Wang, P.~Cieplak, and P.~A. Kollman,
\newblock J. Comput. Chem. {\bf 21}, 1049 (2000).

\bibitem{GAFFcite}
J.~Wang, R.~M. Wolf, J.~W. Caldwell, P.~A. Kollman, and D.~A. Case,
\newblock J. Comput. Chem. {\bf 25}, 1157 (2004).

\bibitem{FMOMD1}
Y.~Komeiji, T.~Nakano, K.~Fukuzawa, Y.~Ueno, Y.~Inadomi, T.~Nemoto,
  M.~Uebayasi, D.~G. Fedorov, and K.~Kitaura,
\newblock Chem. Phys. Lett. {\bf 372}, 342 (2003).

\bibitem{FMOsawada1}
T.~Sawada, D.~G. Fedorov, and K.~Kitaura,
\newblock J. Am. Chem. Soc. {\bf 132}, 16862 (2010).

\bibitem{ionl1}
E.~I. Izgorodina, J.~Rigby, and D.~R. MacFarlane,
\newblock Chem. Commun. {\bf 48}, 1493 (2012).

\bibitem{ionl2}
J.~Rigby, S.~B. Acevedo, and E.~I. Izgorodina,
\newblock J. Chem. Theory Comput. {\bf 11}, 3610 (2015).

\bibitem{ionl3}
E.~I. Izgorodina, Z.~L. Seeger, D.~L.~A. Scarborough, and S.~Y.~S. Tan,
\newblock Chem. Rev. {\bf 117}, 6696 (2017).

\bibitem{enzyme}
Y.~Abe, M.~Shoji, Y.~Nishiya, H.~Aiba, T.~Kishimoto, and K.~Kitaura,
\newblock Phys. Chem. Chem. Phys. {\bf 19}, 9811 (2017).

\bibitem{1L2yvib}
Z.~Ahmed, I.~A. Beta, A.~V. Mikhonin, and S.~A. Asher,
\newblock J. Am. Chem. Soc. {\bf 127}, 10943 (2005).

\end{thebibliography}
   
\newpage
\clearpage
\clearpage

TABLE captions.
\begin{table}[h!]
\caption[]{
   Accuracy of the approximate and analytic gradient of FMO based QM/MM,
evaluated in comparison to numerical gradient (in hartree/bohr)
 for the formate+DMEDAH ionic liquid.
\label{TABLE01}}
\begin{tabular}{lrr}\hline
                &   approximate$^a$ &  exact  \\\hline
 Maximum error  &    0.00027171      &  0.00000368     \\
 RMSE           &    0.00006088      &  0.00000065     \\\hline
\end{tabular}
\\
$^a$ Neglecting the response terms in eq \ref{ResponseTerm}.
\end{table}%

\clearpage
\clearpage

\begin{table}[h!]
\caption[]{
   Frequencies (cm$^{-1}$) and intensities (Debye$^2$/(mass \AA$^2$) for
prominent IR peaks in the solvated Trp-cage protein, computed
   with FMO/FDD
\label{TABLE02}}
\begin{tabular}{lrrrr}\hline
                       &  \multicolumn{4}{c}{water B3LYP-D}                                \\\hline
   mode                &   \multicolumn{2}{c}{FMO3/MM$^a$} & \multicolumn{2}{c}{QM/MM$^a$} \\\hline
   mode                &   frequency  &  intensity         & frequency  & intensity   \\\hline
   sym H-O-H bending   &    1752      &    4.23            &  1752      &   4.27      \\
   sym O-H stretch     &    3680      &    7.49            &  3680      &   7.40      \\
  asym O-H stretch     &    3792      &    3.72            &  3792      &   3.66      \\\hline
                       &  \multicolumn{4}{c}{1l2y HF-D}                          \\\hline
   mode                &   \multicolumn{2}{c}{QM/MM$^b$} & \multicolumn{2}{c}{all QM$^c$}  \\\hline
   mode                &   frequency  &  intensity    & frequency  & intensity   \\\hline
   Amide III           &   1339       &  3.0          &  1354    &  3.4        \\
   Amide II            &   1740       &  6.8          &  1745    &  7.3        \\
   Amide I             &   1914       & 10.6          &  1939    & 10.2        \\      
   N-H stretch         &   3708       & 11.3          &  3707    & 11.9        \\\hline
\end{tabular}
\\
$^a$ 9 and 365 water molecules are treated with QM and MM, respectively.
$^b$ 368 and 1604 water molecules are  treated with QM and MM, respectively.
$^c$ 1972 water molecules are treated with QM.
\end{table}%

\clearpage
\clearpage

\begin{table}[h!]
\caption[]{
Maximum difference (Max) and RMSD of vibrational frequencies (cm$^{-1}$)
    taking the (envir=MM) results for the QM size $R$ (\AA) of 20 \AA\ as the reference.
\label{TABLE03}}
\begin{tabular}{rrrrr}\hline
                &   \multicolumn{2}{c}{envir=MM} & \multicolumn{2}{c}{envir=none} \\\hline
     $R$                &   Max &  RMSD        & Max & RMSD        \\\hline
               \multicolumn{5}{c}{formate in ionic liquid}                                      \\\hline
     20                &    0.0     &   0.0      &    1.2   &    0.8    \\
     18                &    0.5     &   0.3      &    6.5   &    3.3    \\
     15                &    2.2     &   1.2      &   10.6   &    5.0    \\      
     12                &    0.8     &   0.4      &    4.5   &    2.2    \\      
     10                &    2.0     &   1.0      &    9.6   &    4.6    \\      
      8                &    4.1     &   2.0      &    3.6   &    2.0    \\      
      7                &    2.8     &   1.3      &   33.8   &   17.4    \\      
      6                &    5.0     &   3.2      &    9.7   &    4.8    \\\hline
               \multicolumn{5}{c}{DMEDAH in ionic liquid}                               \\\hline
     20                &     0.0    &    0.0     &    7.8   &    4.1    \\
     18                &     2.4    &    1.1     &   12.1   &    6.7    \\
     15                &     1.7    &    1.1     &   17.4   &    9.7    \\      
     12                &     1.9    &    1.4     &   31.8   &   18.0    \\      
     10                &     7.0    &    4.4     &   38.2   &   26.5    \\      
      8                &     7.5    &    4.6     &   42.7   &   20.0    \\      
      7                &    23.4    &   14.0     &   33.6   &   16.2    \\      
      6                &    65.0    &   37.2     &   43.3   &   32.8    \\\hline
               \multicolumn{5}{c}{Trp-6 in Trp-cage}                                         \\\hline
     20                &     0.0    &   0.0      &   30.6   &   15.5    \\
     18                &     6.4    &   4.7      &   34.6   &   19.1    \\
     15                &     1.7    &   1.1      &   33.2   &   16.4    \\      
     12                &    18.2    &   9.1      &   30.4   &   15.6    \\      
     10                &    65.2    &  36.1      &   37.6   &   29.6    \\      
      8                &    51.2    &  36.4      &   54.4   &   40.9    \\\hline
\end{tabular}
\\
\end{table}%

\clearpage
\clearpage

\begin{table}[h!]
\caption[]{
    Maximum difference (Max) and RMSD of vibrational intensities (\%)
    taking the (envir=MM) results for the QM size $R$ (\AA) of 20 \AA\ as the reference.
\label{TABLE02a}}
\begin{tabular}{rrrrr}\hline
                &   \multicolumn{2}{c}{envir=MM} & \multicolumn{2}{c}{envir=none} \\\hline
     $R$                &   Max &  RMSD        & Max & RMSD        \\\hline
               \multicolumn{5}{c}{formate in ionic liquid}                                      \\\hline
     20                &     0.0    &     0.0    &     5.6  &     2.9   \\
     18                &     1.0    &     0.5    &    11.5  &     6.0   \\
     15                &     3.9    &     2.1    &    17.1  &     9.1   \\
     12                &     1.2    &     0.8    &     7.0  &     4.2   \\      
     10                &     3.4    &     1.9    &    16.5  &     7.9   \\      
      8                &     6.3    &     3.2    &     7.0  &     4.8   \\
      7                &     8.0    &     6.0    &    62.2  &    33.1   \\
      6                &    15.3    &    10.7    &    12.1  &     8.9   \\\hline
               \multicolumn{5}{c}{DMEDAH in ionic liquid}                                       \\\hline
     20                &      0.0   &    0.0     &   16.5   &    7.5    \\
     18                &      3.1   &    1.7     &   11.0   &    6.1    \\
     15                &      2.7   &    1.8     &   24.0   &   12.2    \\      
     12                &      3.7   &    2.8     &   32.6   &   15.4    \\      
     10                &     15.4   &    9.9     &   36.8   &   24.8    \\      
      8                &     17.0   &   10.4     &   43.2   &   24.1    \\      
      7                &     31.2   &   19.6     &   41.0   &   25.5    \\      
      6                &     94.8   &   63.3     &   95.5   &   69.4    \\\hline
               \multicolumn{5}{c}{Trp-6 in Trp-cage  }                                       \\\hline
     20                &      0.0   &    0.0      &   4.7    &    4.0     \\
     18                &      2.0   &    1.7      &   6.2    &    4.8     \\
     15                &      1.1   &    0.9      &   4.3    &    4.0     \\    
     12                &      7.2   &    4.3      &   7.3    &    5.1     \\   
     10                &      9.8   &    5.9      &  15.8    &    9.7     \\    
      8                &     53.8   &   33.2      &  19.9    &   12.0     \\\hline
\end{tabular}
\\
\end{table}%

\clearpage
\clearpage

\begin{table}[h!]
\caption[]{
   Vibrational frequencies (cm$^{-1}$) in Trp-cage using the experimental notation w$i$.\cite{1L2yvib}
\label{TABLE02b}}
\begin{tabular}{llrrrrr}\hline
method  & environment               &      w3       &     w8     &    w7     &    w16   &    w18       \\\hline
HF  & none$^a$                      &   1530        &   1492     &  1415     &  995     &   740        \\
FMO-HF & protein$^a$                &   1540        &   1442     &  1358     &  993     &   737        \\
FMO-HF/MM & protein+water$^a$       &   1567        &   1440     &  1357     &  1021    &   742        \\
FMO-HF/MM & protein+water$^b$       &   1554        &   1449     &  1351     &  1018    &   748        \\
expt\cite{1L2yvib} & protein+water  &   1558        &   1450     &  1365     &  1014    &   765        \\\hline
\end{tabular}
\\
$^a$ For a single minimum 
$^b$ Combining results from 100 conformations.
\end{table}%

\clearpage
\section*{Figure captions}

\clearpage


\newpage

\end{document}